\documentclass[12pt]{article}
\usepackage{amsfonts,amsmath,amssymb}
\usepackage{graphicx}

\newfont{\twelvemsb}{msbm10 scaled\magstep1}
\newfont{\eightmsb}{msbm8}
\newfam\msbfam
\textfont\msbfam=\twelvemsb \scriptfont\msbfam=\eightmsb
\catcode`\@=11
\def\Bbb{\ifmmode\let\next\Bbb@\else
\def\next{\errmessage{Use \string\Bbb\space only in math mode}}\fi\next}
\def\Bbb@#1{{\fam\msbfam{{#1}}}}

\newcommand{\be}{\begin{equation}}
\newcommand{\ee}{\end{equation}}
\newcommand{\ba}{\begin{eqnarray}}
\newcommand{\ea}{\end{eqnarray}}

\newcommand{\m}{\mathcal}

\newcommand{\nn}{\nonumber}
\newcommand{\q}{\theta}

\begin{document}

\sloppy
\renewcommand{\thefootnote}{\fnsymbol{footnote}}
\newpage
\setcounter{page}{1} \vspace{0.7cm}
\begin{flushright}
30/01/17
\end{flushright}
\vspace*{1cm}
\begin{center}
{\bf The contribution of scalars to ${\cal N}=4$ SYM amplitudes} \\
\vspace{.6cm} {Alfredo Bonini$^{a,b}$, Davide Fioravanti $^a$, Simone Piscaglia $^{c}$, Marco Rossi $^d$}
\footnote{E-mail: bonini@bo.infn.it, fioravanti@bo.infn.it, piscaglia@th.phys.titech.ac.jp
, rossi@cs.infn.it} \\
\vspace{.3cm} $^a$ {\em Sezione INFN di Bologna, Dipartimento di Fisica e Astronomia,
Universit\`a di Bologna} \\
{\em Via Irnerio 46, 40126 Bologna, Italy}\\

\vspace{.3cm} $^b$ {\em Institut de Physique Th\'eorique, CEA, 
Saclay, F-91191 Gif-sur-Yvette, France}\\

\vspace{.3cm} $^c$ {\em Department of Physics,
Tokyo Institute of Technology,
Tokyo 152-8551, Japan} 

\vspace{.3cm} $^d${\em Dipartimento di Fisica dell'Universit\`a
della Calabria and INFN, Gruppo collegato di Cosenza} \\
{\em Arcavacata di Rende, Cosenza, Italy} \\
\end{center}
\renewcommand{\thefootnote}{\arabic{footnote}}
\setcounter{footnote}{0}
\begin{abstract}
{\noindent We analyse the contribution of $2n$ scalars to the Operator Product Expansion series for MHV gluon scattering amplitudes/polygonal Wilson loops in planar ${\cal N}=4$ SYM. Hence, we sum up on $2n$ for large coupling $\lambda$: the logarithm of the amplitude is proportional to $\sqrt{\lambda}$, thus competing, unexpectedly,  with the famous classical contribution. We give explicit expressions for the first terms at large $\lambda$ in case of two and four scalars. For finalising this analysis, we find profitable an explicit computation of the $2n$-scalar term at any coupling by means of Young tableaux, paralleling, under certain aspects, the Nekrasov-Okounkov expressions for $\mathcal N =2$ SYM prepotential.}

\end{abstract}
\vspace{6cm}

\newpage



Remarkably, scattering amplitudes between gluons in planar ($N_c \rightarrow \infty$) ${\cal N}=4$ SYM have recently been conjectured to be described by the same string as (expectation values of) null polygonal Wilson loops (Wls), and thus behave at strong coupling as an exponential of the square root of the 't Hooft coupling $\lambda=N_c g_{YM}^2$ ({\it i.e.} the classical string minimal area) \cite{AM-amp}. On the other hand, pursuing the idea of a non-local Operator Product Expansion (OPE) for the polygon when two adjacent sides become aligned \cite{Anope}, this Wl has been importantly given an expression in terms of an exact series over all GKP excitations \cite{BSV1,BSV4}. And this is the multi-particle Form Factor (FF) series ({\it cf.} for instance \cite{Smirnov} and references therein\footnote{Although in the present non-relativistic case some of the FF properties change significantly.}) for a specific {\it twist} field $P$ of the integrable theory with non-relativistic scattering matrix on the GKP vacuum \cite{FPR1,Basso:2013pxa,FPR2}: although simple, this observation is crucial for the following strong coupling analysis. In fact, the square root or, more precisely, the exact minimal area is furnished by a Thermodynamic Bethe Ansatz (TBA) \cite{TBuA}, which, in its turn, is exactly reproduced by the OPE series contributions of gluons and mesons only \cite{FPR2}: all the other particles, including scalars, should, {\it naively}, intervene at sub-leading order \cite{BSV1,BSV4,FPR2,BFPR}. Yet, \cite{BSV4} made the puzzling proposal that the strong coupling contribution of scalars should be of the same order ($\sqrt\lambda$) as the semiclassical prediction: in fact, the low energy limit of string theory is given by an O(6) non-linear sigma model (nlsm) \cite{AM}, and the latter shows in the PP correlator the peculiar scaling of a 2D CFT at short distances (as the dynamical mass gap goes to zero with the coupling \cite{AM}). Moreover, this relevant scaling was corroborated by Monte Carlo simulations on the few-particles terms of the series \cite{BSV4} (then refined by \cite{BEL}). Nevertheless, what seems missing so far is an analytical derivation of this simple square root behaviour directly from the OPE series. This also supports the conjectural form \cite{BSV4} of  its multi-particle terms and is the main contribution of this letter. In fact, upon refining a method of FF theory \cite{Smirnov} in case of asymptotically free theories (here the O(6) nlsm), we easily extract, at fixed number of scalars, a $\sqrt\lambda$ in front of any connected term of the logarithm of the series.
In this manner, we find an expression for the whole coefficient in front of $\sqrt\lambda$ and the explicit values of the first terms (proportional to $\sqrt{\lambda}$, $\ln \lambda$ and $\lambda ^0$) of the two (and four) scalar contributions. To this purpose, we use the two-body product for the dynamical part of the FF expression (with simplifications at large coupling) and also find possible manipulations of the coupling-independent matrix part. This does not enjoys a two-body form, but, thanks to its R-symmetry origin, allows us an explicit procedure of computation based on sums over Young tableaux, an interesting resemblance with Nekrasov-Okounkov method for instanton partition functions of $\mathcal N =2$ SYM \cite{Nekrasov}.

\section{Non-perturbative scalars in the Wilson loop}
\label{Wilson loops}

As anticipated, the OPE series is an infinite sum over the asymptotic states of particles (gluons, fermions and scalars) \cite{BSV1} of the GKP scattering theory \cite{FPR1,FPR2}. As such, it is simple to single out which contribution, $W$, the scalars give to the hexagonal\footnote{Generalisation to the other polygons goes along similar lines\cite{35}.} Wilson loop by constraining the sum only on them ($2n$ is their number), and then to consider their non-perturbative strong coupling regime, where they indeed decouple and form a relativistic O(6) nlsm \cite{AM}:
\be
\label{Wilson}
W=\sum_{n=0}^{\infty}W^{(2n)} \, , \quad W^{(2n)}=\frac{1}{(2n)!}\int\prod_{i=1}^{2n}\frac{d\theta_i}{2\pi}\,G^{(2n)}(\theta_1,\cdots,\theta_{2n})\, e^{- z\sum\limits_{i=1}^{2n}\cosh\theta_i} \, ,
\ee
with Bethe rapidities $u_i$, via $\q_i=\frac{\pi}{2}u_i$, and the parameter $z=m_{gap}\sqrt{\tau^2+\sigma^2}$, containing the two conformal ratios $\sigma$, $\tau $, proportional to their dynamically generated mass \cite {AM,FGR},
\be
m_{gap}(\lambda)=\frac{2^{1/4}}{\Gamma (5/4)} \lambda ^{1/8}e^{-\sqrt{\lambda}/4} (1+O(1/\sqrt{\lambda} ) ) \, .
\ee
At any coupling, each function $G^{(2n)}$ factorises into a coupling-dependent dynamical part, $\Pi_{dyn}^{(2n)}$, and a coupling-independent matrix one, $\Pi_{mat}^{(2n)}$, reflecting the structure of scalars under the internal $SO(6)$ symmetry \cite{BSV4}:
\be\label{Gi2n}
G^{(2n)}(u_1,\cdots,u_{2n})=\Pi_{dyn}^{(2n)}(u_1,\cdots,u_{2n})\,\Pi_{mat}^{(2n)}(u_1,\cdots,u_{2n}) \, .
\ee
Interestingly, the dynamical part is a product over a two-particle function, which also becomes relativistic at strong coupling (as already stated $\q_i=\frac{\pi}{2}u_i$, and $\q=\frac{\pi}{2}u$)
\be\label{dynamical}
\Pi_{dyn}^{(2n)}(u_1,\cdots,u_{2n})=\mu^{2n}\prod_{i<j}^{2n}\Pi(u_i-u_j) \, , \quad
\Pi(u)=
\frac{8\theta\tanh \left(\frac{\theta}{2}\right)\Gamma \left (\frac{3}{4}+\frac{i\theta}{2\pi} \right)
\Gamma \left (\frac{3}{4}-\frac{i\theta}{2\pi} \right)}{\pi\Gamma \left (\frac{1}{4}+\frac{i\theta}{2\pi} \right) \Gamma \left (\frac{1}{4}-\frac{i\theta}{2\pi} \right)}\, ,
\quad  \mu=\frac{2\Gamma\left(\frac{3}{4}\right)}{\sqrt{\pi}\Gamma\left(\frac{1}{4}\right)} \, .
\ee
On the contrary, the matrix part does not depend on $\lambda$, but only on differences of rapidities $u_i-u_j$, and has been conjectured to be this involved integral over the isotopic variables \cite{BSV4, BFPR},
\ba\label{Pi_mat}
\Pi_{mat}^{(2n)}(u_1,\dots,u_{2n}) &=& \frac{1}{(2n)!(n!)^2}\int_{-\infty}^{+\infty}
\prod_{k=1}^{n}\frac{da_k}{2\pi}
\prod_{k=1}^{2n}\frac{db_k}{2\pi}\prod_{k=1}^{n}\frac{dc_k}{2\pi} \times \\
&\times& \frac{\displaystyle\prod_{i<j}^{n} g(a_i-a_j) \prod_{i<j}^{2n} g(b_i-b_j) \prod_{i<j}^{n} g(c_i-c_j)}
{\displaystyle \prod_{j=1}^{2n} \left(\prod_{i=1}^{n} f(a_i-b_j) \prod_{k=1}^{n} f(c_k-b_j)
\prod_{l=1}^{2n} f\left (u_l-b_j \right)\right)} \ , \nn
\ea
with $f(x)=x^2+\frac{1}{4} \, , \,\, g(x)=x^2(x^2+1)$. As this form entails our final results, which reproduce CFT considerations of \cite{BSV4}, it will receive strong confirmation. Moreover, it will imply a suggestive alternative sum over Young tableaux in next section.


Now, the basic idea is to evaluate the logarithm of (\ref{Wilson}) by passing to the connected counterparts of the $G^{(2n)}$. With this proposal in mind, we need to study the behaviour of $G^{(2n)}$ when we shift $m$ rapidities by a large amount $\Lambda\rightarrow\infty$, while holding fixed the remaining $2n-m$ ones. We will show that for even $m$ the function $G^{(2n)}$ enjoys the {\it asymptotic factorisation} into the two functions with fewer rapidities: $\Lambda\rightarrow\infty
$\be\label{fact}
G^{(2n)}\rightarrow G^{(m)}\,G^{(2n-m)} \, , \quad 2\leq m\leq 2n-2 \, ;
\ee
on the other hand, if $m$ is odd the function $G^{(2n)}$ goes to zero as $1/\Lambda ^2$. Except for this power-like behaviour, this is what has been shown in \cite{Smirnov}: in fact, we shall adapt those ideas in the case of asymptotic freedom and we will be general for these theories which -- to the best of our knowledge -- have not been treated in this respect. First, the dynamical factor (\ref{dynamical}) (also considered for odd number $m$ of scalars) enjoys the factorisation
\be
\Pi_{dyn}^{(2n)}(u_1+\Lambda,\cdots,u_{m}+\Lambda,u_{m+1},\cdots,u_{2n})
\longrightarrow\Lambda^{2m(2n-m)}\Pi_{dyn}^{(m)}(u_1,\cdots,u_{m})\Pi_{dyn}^{(2n-m)}(u_{m+1},\cdots,u_{2n}) \ ,
\label {dynasy}
\ee
as a consequence of the asymptotic behaviour
$\Pi(u)\simeq u^2$ when $u \rightarrow \infty$. As it is more involved how to deal with the matrix part  (\ref{Pi_mat}), we would rather tackle, first, the simplest non trivial case, \textit{i.e.} $\Pi_{mat}^{(4)}\rightarrow \Pi_{mat}^{(2)}\Pi_{mat}^{(2)}$. When we shift by a large amount $\Lambda$ two rapidities, say $u_1\to u_1+\Lambda ,\,u_2\to u_2+\Lambda$, then the integrals (\ref{Pi_mat}) receive the main contribution from the region in which one auxiliary variable $a$, two $b$, and one $c$ are large. Therefore, we can rewrite (\ref{Pi_mat}), upon shifting by $\Lambda$, for instance, $a_1,b_1,b_2,c_1$ as
\ba
&&\Pi_{mat}^{(4)}(u_1+\Lambda,u_2+\Lambda,u_3,u_4)=\frac{1}{4!4}\int_{-\infty}^{+\infty}\frac{da_1db_1db_2dc_1}{(2\pi)^4}\frac{g(b_1-b_2)}{\displaystyle\prod_{i=1}^2f(a_1-b_i)f(c_1-b_i)\prod_{i,j=1}^{2}f(u_i-b_j)}\times \nn\\
&& \times \int_{-\infty}^{+\infty}\frac{da_2db_3db_4dc_2}{(2\pi)^4}\frac{g(b_3-b_4)}{\displaystyle\prod_{i=3}^4f(a_2-b_i)f(c_2-b_i)\prod_{i,j=3}^{4}f(u_i-b_j)}
\,\m{R}^{(4,2)}(a_1,a_2,b_1,\dots ,b_4,c_1,c_2;\Lambda) \, ,\nn\\ \label {Pi_mat2}
\ea
with
\ba
&& \m{R}^{(4,2)}(a_1,a_2,b_1,\dots ,b_4,c_1,c_2;\Lambda) =
\frac{\displaystyle\prod_{i=1}^2\prod_{j=3}^{4}g(b_i-b_j+\Lambda)}{\displaystyle\prod_{i=1}^{2}\prod_{j=3}^{4}f(u_i-b_j+\Lambda)f(u_j-b_i-\Lambda)} \times\nn\\
&& \times \frac{g(a_1-a_2+\Lambda)g(c_1-c_2+\Lambda)} {\displaystyle\prod_{i=3}^{4}f(a_1-b_i+\Lambda)f(c_1-b_i+\Lambda)\prod_{i=1}^{2}f(a_2-b_i-\Lambda)f(c_2-b_i-\Lambda)} \ .
\ea
Now we can effectively perform the limit: in consideration of the expansion
\be
\m{R}^{(4,2)}(a_1,a_2,b_1,\dots ,b_4,c_1,c_2;\Lambda) =
\Lambda^{-8}\left[1+O\left(\frac{1}{\Lambda}\right)\right] \, , \quad \Lambda \rightarrow +\infty \, ,
\ee
and of all the possible exchanges of auxiliary variables (of the same type), {\it i.e.} a multiplicity factor $24$, we obtain
\ba
\Pi_{mat}^{(4)}(u_1+\Lambda,u_2+\Lambda,u_3,u_4)\simeq 24\Lambda^{-8}\frac{1}{4!4}\int\frac{da_1db_1db_2dc_1}{(2\pi)^4}\frac{g(b_1-b_2)}{\displaystyle\prod_{i=1}^2f(a_1-b_i)f(c_1-b_i)\prod_{i,j=1}^{2}f(u_i-b_j)}\times \nn\\
\times \int\frac{da_2db_3db_4dc_2}{(2\pi)^4}\frac{g(b_3-b_4)}{\displaystyle\prod_{i=3}^4f(a_2-b_i)f(c_2-b_i)\prod_{i,j=3}^{4}f(u_i-b_j)}=\Lambda^{-8}\Pi_{mat}^{(2)}(u_1,u_2)\Pi_{mat}^{(2)}(u_3,u_4) \ . \nn\\
\ea
Assembling together (\ref{Gi2n}), the four scalar asymptotic factorisation is then proven:
\be
G^{(4)}(u_1+\Lambda, u_2+\Lambda, u_3,u _4)\ \overset{\Lambda\rightarrow\infty}{\longrightarrow}\ G^{(2)}(u_1,u_2)G^{(2)}(u_3,u_4) + O(\Lambda^{-2})\, .
\ee
The $O(1/\Lambda)$ correction vanishes thanks to a refined cancellation coming from the matrix part ($\m{R}^{(4,2)}$) and the dynamical one: this fact will be relevant in the following. We do not need to change the scheme in the most general case $u_i\to u_i+\Lambda$ for $1\leq i\leq m$, but only to separate odd $m=2k-1$ from even $m=2k$. In a unified manner, we can describe the shifts $a_j\to a_j+\Lambda$ and $c_j\to c_j+\Lambda$ for $1\leq j\leq k$, along with $b_i\to b_i+\Lambda$ for $1\leq i\leq m$, namely
\ba \label{Intpimat}
&&\Pi_{mat}^{(2n)}(u_1+\Lambda,\cdots,u_{m}+\Lambda,u_{m+1},\cdots,u_{2n})= \\
&& = \frac{1}{(2n)!(n!)^2}\int\prod_{i=1}^k\frac{da_i dc_i}{(2\pi)^2}\prod_{i=1}^m\frac{db_i}{2\pi}\,
\frac{\displaystyle\prod_{i<j,\,i=1}^{k}\left[g(a_i-a_j)g(c_i-c_j)\right]\displaystyle\prod_{i<j,\,i=1}^{m}g(b_i-b_j)}
{\displaystyle\prod_{j=1}^{m}\left[\prod_{i=1}^{k}f(a_i-b_j)f(c_i-b_j)\prod_{l=1}^{m}f(u_l-b_j)\right]}\times \nn\\
&& \times \int\prod_{i=k+1}^n\frac{da_i dc_i}{(2\pi)^2}\prod_{i=m+1}^{2n}\frac{db_i}{2\pi}\,
\frac{\displaystyle\prod_{i<j,\,i=k+1}^{n}\left[g(a_i-a_j)g(c_i-c_j)\right]\displaystyle\prod_{i<j,\,i=m+1}^{2n}g(b_i-b_j)}
{\displaystyle\prod_{j=m+1}^{2n}\left[\prod_{i=k+1}^{n}f(a_i-b_j)f(c_i-b_j)\prod_{l=m+1}^{2n}f(u_l-b_j)\right]}
\,\m{R}^{(2n,m)}(a_1,\dots , c_{2n};\Lambda) \,\, , \nn
\ea
with
\ba
&& \m{R}^{(2n,m)}(a_1,\dots , c_{2n};\Lambda)= \frac{\displaystyle\prod_{i=1}^{m}\prod_{j=m+1}^{2n}g(b_i-b_j+\Lambda)}{\displaystyle\prod_{i=1}^{m}\prod_{j=m+1}^{2n}f(u_j-b_i-\Lambda)f(u_i-b_j+\Lambda)} \times \nn\\
&& \times
\frac{\displaystyle\prod_{i=1}^{k}\prod_{j=k+1}^{n}g(a_i-a_j+\Lambda)g(c_i-c_j+\Lambda)}
{\displaystyle\prod_{j=1}^{m}\prod_{i=k+1}^{n}f(a_i-b_j-\Lambda)f(c_i-b_j-\Lambda)
\prod_{j=m+1}^{2n}\prod_{i=1}^{k}f(a_i-b_j+\Lambda)f(c_i-b_j+\Lambda)} \ .
\ea
In general $\m{R}^{(2n,m)} \simeq \Lambda^{4(n-k)(2k-m)}\Lambda^{-4k(2n-m)}$, hence the following factorisation of the matrix part (\ref {Intpimat}), with even $m=2k$,
\be
\Pi_{mat}^{(2n)}(u_1+\Lambda,\cdots,u_{2k}+\Lambda,u_{2k+1},\cdots,u_{2n})\longrightarrow  \Lambda^{-2m(2n-m)}\Pi_{mat}^{(2k)}(u_1,\cdots,u_{2k})\Pi_{mat}^{(2n-2k)}(u_{2k+1},\cdot\cdot,u_{2n}) \, 
\label{pimat-fact}
\ee
can be put together with (\ref{dynasy}). This entails $G^{(2n)}(u_1+\Lambda,\cdots,u_{m}+\Lambda,u_{m+1},\cdots,u_{2n})$ weighted by a factor $\Lambda^{-2(m-2k)^2}$, which means, for odd $m=2k-1$, the suppression $G^{(2n)}\simeq \Lambda^{-2}$. Furthermore, for even $m=2k$, the asymptotic factorisation (\ref{fact}) is eventually achieved:
\be\label{factorization}
G^{(2n)}(u_1+\Lambda,\cdots,u_{2k}+\Lambda,u_{2k+1},\cdots,u_{2n})\ \overset{\Lambda\to\infty}{\longrightarrow}\ G^{(2k)}(u_1,\cdots,u_{2k})\,G^{(2n-2k)}(u_{2k+1},\cdots,u_{2n}) + O(\Lambda^{-2})\ .
\ee
Subtly, for obtaining this formula we had to consider all the possible shifts of the auxiliary variables (for given $2k$ and $2(n-k)$ particle rapidities) within the integrand (\ref{Intpimat}), and produced a multiplicity factor ${n \choose k}^2{2n \choose 2k}$, which, once combined with the present factorials, as $\frac{1}{(2n)!(n!)^2}{n \choose k}^2{2n \choose 2k}=\frac{1}{(2k)!(k!)^2}\frac{1}{(2n-2k)!((n-k)!)^2}$, reproduces the correct factorials of $G^{(2k)}$ and $G^{(2n-2k)}$.
Actually, we need, in addition to (\ref{factorization}), a guessable extension of it with different large shifts $\Lambda_i$ (we will prove it in details \cite{35}): for this new necessity is a consequence of the power like correction (physically ascribable to asymptotic freedom) in place of the exponential one of \cite{Smirnov}. Now, we can profitably pass to the $2n$ connected functions $g^{(2n)}$
\be\label{logW}
{\cal F}=\ln W = \sum_{n=1}^{\infty}{\cal F}^{(2n)}=\sum_{n=1}^{\infty}\frac{1}{(2n)!}\int\prod_{i=1}^{2n}\frac{d\theta_i}{2\pi}g^{(2n)}(\theta_1,\cdots,\theta_{2n})e^{-z\sum_{i=1}^{2n}\cosh\theta_i} \, .
\ee
A well-known combinatorial definition tells us that the original multi-particle functions $G^{(2n)}$ are expressed in terms of the connected ones, $g^{(2k)}$, as a sum over all the possible ways, $\left\{m_k\right\}$, to arrange $2n$ particles in subgroups of (even) particles\footnote{A similar formula holds without the parity constraint.}:
\be
G^{(2n)}=\sum_{\left\{m_k\right\}}\sum_{\textit{d.e.}}\displaystyle\prod_{k=1}^{n}\underbrace{g^{(2k)}\cdots g^{(2k)}\,}_\text{$m_k$ terms} \, ,
\label{Gg}
\ee
where $m_k$, $k=1,\cdots,n$, is the number of $g^{(2k)}$, depending on different rapidities (thus with the constraint $\sum_{k=1}^{n}2km_k=2n$) and any {\it different exchange} (d.e.) is an exchange between two rapidities belonging to two {\it different} $g^{(2k)}$ and producing a {\it different} term. Their number is
$\left(\prod_{k=1}^{n}1/m_k!\right) (2n)!/\prod_{k=1}^{n}((2k)!)^{m_k}$ . First examples are $G_{12}^{(2)}=g_{12}^{(2)}$, $G_{1234}^{(4)}=g_{1234}^{(4)}+g_{12}^{(2)}g_{34}^{(2)}+g_{13}^{(2)}g_{24}^{(2)}+g_{14}^{(2)}g_{23}^{(2)}= g_{1234}^{(4)} + (g_{12}^{(2)}g_{34}^{(2)} + 2 \textit{ d.e.})$, $G_{123456}^{(6)}=g_{123456}^{(6)} + (g_{12}^{(2)}g_{3456}^{(4)} + 14\textit{ d.e.}) + (g_{12}^{(2)}g_{34}^{(2)}g_{56}^{(2)} +14 \textit{ d.e.})$. Conversely, the first examples of connected functions $g^{(2m)}$ in terms of $G^{(2n)}$ are: $g_{12}^{(2)}=G_{12}^{(2)}$, $g_{1234}^{(4)}=G_{1234}^{(4)}-G_{12}^{(2)}G_{34}^{(2)}-G_{13}^{(2)}G_{24}^{(2)}-G_{14}^{(2)}G_{23}^{(2)} = G_{1234}^{(4)}-(G_{12}^{(2)}G_{34}^{(2)} + 2 \textit{ d.e.})$, $g_{123456}^{(6)}=G_{123456}^{(6)} - (G_{12}^{(2)}G_{3456}^{(4)} + 14\textit{ d.e.}) + 2(G_{12}^{(2)}G_{34}^{(2)}G_{56}^{(2)} +14 \textit{ d.e.})$. These formul{\ae} have no difference w.r.t. the previous ones, but for the presence of signs and an additional factor: this is a general feature for the inverse expression \cite{35}
\be
g^{(2n)}=\sum_{\left\{m_k\right\}}f(\left\{m_k\right\})\sum_{\textit{d.e.}}\displaystyle\prod_{k=1}^{n}\underbrace{G^{(2k)}\cdots G^{(2k)}\,}_\text{$m_k$ terms} \, ,
\label{gG}
\ee
where every product carries the factor $f(\left\{m_k\right\})=(-1)^{p}p!$, $p=\sum_{k=1}^{n}m_k - 1$.
Now, a crucial observation enters the stage: factorisation (\ref{factorization}) implies that the connected functions vanish (rapidly enough to ensure their integrability, see below) whenever a subset of rapidities is sent far away from all the others by a great quantity $\Lambda$, \textit{i.e.}
\be
\lim_{\Lambda\to\infty}g^{(2n)}(\theta_1+\Lambda,\cdots,\theta_{m}+\Lambda,\theta_{m+1},\cdots,\theta_{2n})\simeq \frac{1}{\Lambda ^2} \rightarrow 0 \, , \qquad\mbox{for}\ \ m<2n \, . \label {g-asy}
\ee
This follows from the specific combinatorial form of the r.h.s. of (\ref{gG}) once manipulated by means of the factorisation (\ref{factorization}) to give rise to peculiar cancellations. Conversely, (\ref {g-asy}) entails the factorisation (\ref{factorization})  via (\ref{Gg}),
thus establishing the equivalence of the two properties.
Clearly, the limit (\ref {g-asy}) decides the small $z$ behaviour of the logarithm of the Wilson loop, and, in particular, its power-like decay (due to asymptotic freedom) implies the presence of a $\ln (\ln 1/z)$ term. Actually, we need again (and can prove \cite{35}), an extension of (\ref{g-asy}) with different large shifts $\Lambda_i$.\\
To derive the conformal limit at small $z$, we shall consider the multi-integral $I^{(2n)}= (2n)! (2\pi)^n{\cal F}^{(2n)}$ in the generic term of the series (\ref {logW}), and integrate on the connected function  $g^{(2n)}$ which depends only on the differences $\alpha _i=\theta _{i+1} -\theta _1$, $i=1,\ldots , 2n-1$. Thus, upon isolating the integration on $\theta _1$
\be
I^{(2n)}=\int d\theta _1 \prod _{i=1}^{2n-1}d \alpha _i \exp \Bigl [ -z \cosh \theta _1 -z \sum _{i=2}^{2n} \left ( \cosh \theta _1 \cosh \alpha _{i-1} + \sinh \theta _1 \sinh \alpha _{i-1} \right )  \Bigr ]  g^{(2n)}(\alpha _1, \ldots , \alpha _{2n-1}) \, ,
\ee
we notice the convenient definitions $a=1+\sum_{i=2}^{2n} \cosh  \alpha _{i-1}=\xi \cosh \eta$ and $b= \sum_{i=2}^{2n} \sinh  \alpha _{i-1}=\xi \sinh \eta$ for some real $\eta$ (depending on the $\alpha _i$, but not on $\theta _1$), because of the identity $a^2-b^2=2n+2  \sum _{i=2}^{2n} \cosh  \alpha _{i-1} + 2 \sum _{i=2}^{2n} \sum _{j=i+1}^{2n} \cosh (\alpha _{i-1} - \alpha _{j-1})= \xi ^2 >0$. As a consequence
\small
\be
I^{(2n)}= \int \prod _{i=1}^{2n-1}d \alpha _i g^{(2n)}(\alpha _1, \ldots , \alpha _{2n-1})  \int d\theta _1 \exp \Bigl [ -z \xi \cosh (\theta _1 + \eta )  \Bigr ] = 2 \int \prod _{i=1}^{2n-1}d \alpha _i g^{(2n)}(\alpha _1, \ldots , \alpha _{2n-1})  K_0 (z \xi) \, .
\ee
\normalsize
We would be tempted to expand (inside) for small argument $K_0(z\xi)=-\ln z -\ln\xi + \ln 2-\gamma + O(z^2\ln z)$, with $\gamma=0.5772...$ the Euler-Mascheroni constant. Subtly, because of the weak decay (\ref{g-asy}), we need to restrict the integral to the region in which the argument is small and then expand
\be\label{lead-subl}
I^{(2n)}= -2 \ln z  \int \prod _{i=1}^{2n-1}d \alpha _i g^{(2n)}(\alpha _1, \ldots , \alpha _{2n-1}) -2\int_{z\xi<1} \prod _{i=1}^{2n-1}d \alpha _i g^{(2n)}(\alpha _1, \ldots , \alpha _{2n-1})\ln\xi + O(1)\, ,
\ee
where we remove the cut-off only in the first integral since the function $g^{(2n)}$ is integrable, whilst the second one diverges (as $\ln\ln(1/z)$). This mechanism will be even clearer when we analyse later the two particles case ($g^{(2)}(\theta) \sim \theta^{-2}$). For $n>1$ factorisation (\ref{factorization}) allowed us with (\ref{g-asy}) to prove that the function $g^{(2n)}(\alpha _1, \ldots , \alpha _{2n-1})$ goes to zero when some of the rapidity differences $\alpha _i$ are large, but to be integrable we also need a stronger decay when the rapidities go to infinity in different ways. This is an involved matter which also determines the sub-leading behaviour and deserves extended evidence in a longer paper \cite{35}: here we will analyse it up to four particles.
Eventually, the large coupling expansion for the contribution of scalars to the polygonal Wl/MHV scattering amplitude (\ref {logW}) can be systematically set down and gives at first order
\be
\ln W = \frac{\sqrt{\lambda}}{4\pi} \sum _{n=1}^{+\infty} \frac{1}{(2n)!}\int \prod _{i=1}^{2n-1} \frac{d\alpha _i}{2\pi}
g^{(2n)}(\alpha _1, \ldots , \alpha _{2n-1}) + O(\ln\sqrt{\lambda}) \, , \label {fin-scal}
\ee
where we remembered that $\ln(1/z) \sim - \ln m_{gap} \sim \frac{\sqrt{\lambda}}{4}$. In fact, the (divergent) sub-leading term in (\ref{lead-subl}) can be obtained, although without a closed formula because of the new presence of a necessary cut-off w.r.t. \cite{Smirnov}. The latter gives rise to the particular $(\ln 1/z)^s$ factor in the two point 2D CFT correlation function \cite{BSV4}.\\
Nevertheless, our procedure is very effective for computing the analytic expressions of the three leading terms in case of two scalars (improving the Monte Carlo findings of \cite {BSV4} and \cite{BEL}\footnote{The latter gives the (two scalars) leading term an analytical expression.}). In fact, we shall simply specialise (\ref{Pi_mat}) and (\ref {dynamical})
\be
{\cal F}^{(2)}=W^{(2)}=\frac{3\pi}{16}\frac{\Gamma^2(\frac{3}{4})}{\Gamma^2(\frac{1}{4})}\int_{-\infty}^{+\infty}d\theta_1\int_{-\infty}^{+\infty}d\theta_2
\frac{\Pi(\theta_1-\theta_2)}{((\theta_1-\theta_2)^2+\frac{\pi^2}{4})((\theta_1-\theta_2)^2+\pi^2)}e^{-z\cosh\theta_1 -z\cosh\theta_2} \, ,
\ee
change variables $x=\theta _1+\theta _2$, $\theta =\theta _1-\theta _2$ and integrate on $x$,
so to obtain
\be
{\cal F}^{(2)}=\int_{0}^{+\infty}d\theta h(\theta)K_0 \left (2z \cosh \frac{\theta}{2} \right ) , \quad h(\theta)= \frac{3\pi}{4}\frac{\Gamma^2(\frac{3}{4})}{\Gamma^2(\frac{1}{4})}\frac{\Pi(\theta)}{(\theta^2+\frac{\pi^2}{4})(\theta^2+\pi^2)} \, .
\ee
As depicted before in general, we must use the large cut-off $2\ln(1/ z)$:
\be
{\cal F}^{(2)}=\int _{0}^{-2\ln z} d\theta h(\theta) K_0\left(2z \cosh \frac{\theta}{2}\right)+ \int _{-2\ln z}^{+\infty} d\theta h(\theta) K_0\left(2z \cosh \frac{\theta}{2}\right) = I_1+I_2
\ee
Of course, $I_2 \rightarrow 0$ when $z\rightarrow 0$. Then, as illustrated in general above, we expand $K_0(2z \cosh \frac{\theta}{2})$
\be\label{leading}
I_1= -J^{(2)} \ln z - \int _{0}^{-2\ln z} d\theta h(\theta) \ln \left (\cosh \frac{\theta}{2} \right ) - J^{(2)}\gamma + \ln z \int _{-2\ln z}^{\infty} d\theta h(\theta) + O(1/\ln z) \, ,
\ee
with $J^{(2)}=\int_{0}^{\infty} d\theta h(\theta)$. The first contribution is the leading term, proportional to $-\ln z$, while the fourth term yields a finite contribution $-C/2$ with $C=3\Gamma^2(3/4)/(\pi\Gamma^2(1/4))$ because of the asymptotic expansion $h(\theta)=C\theta^{-2}+O(\theta ^{-4})$ at large $\theta$. Which also entails that the second term still contains a (subleading) divergence as it decomposes into
\be
- \int _{0}^{1} d\theta h(\theta) \ln \left (\cosh \frac{\theta}{2} \right ) - \int_{1}^{-2\ln z} d\theta\left[ h(\theta) \ln \left (\cosh \frac{\theta}{2} \right ) -\frac{C}{2\theta}\right]-\frac{C}{2}\int_{1}^{-2\ln z}\frac{d\theta}{\theta} \, .
\ee
For the second addendum above is finite (for $-2\ln z\rightarrow\infty$), but the third one shows a peculiar double log divergence. Eventually, we can argue the general form of
\be
{\cal F}=J \ln (1/z)+s \ln \ln (1/z) + t + O(1/\ln z) \, ,
\label{F}
\ee
with these estimates for the two scalar case ${\cal F}^{(2)}$: $J^{(2)} = 0.03109...$, $s^{(2)}=-C/2 = -0.05454.....$ and 
\small
\be
t^{(2)}=- J^{(2)}\gamma - \frac{C}{2}\left(1+\ln 2\right) - \int_{0}^{1}d\theta h(\theta) \ln \left (\cosh \frac{\theta}{2} \right ) + \int_{1}^{\infty}d\theta \left[\frac{C}{2\theta}-h(\theta) \ln \left ( \cosh \frac{\theta}{2} \right ) \right ]\simeq -0.008 . \nn
\ee
\normalsize
Furthermore, it is not difficult to evaluate the correction coming from the explicit expression of the four scalar connected function $g^{(4)}$ \cite{35}. For instance, we simply integrate it with \texttt{Mathematica}\textsuperscript{\textregistered} and, according to (\ref{fin-scal}), obtain a correction to $J$ of (\ref{F}) by an amount $\delta J= (-3.44\pm 0.01)\cdot 10^{-3}$, {\it i.e.} $J^{(4)}\simeq 0.02765$: this value differs from the 2D-CFT prediction $J=\frac{1}{36}=0.02\bar{7}$ \cite{BSV4,BEL} by just $0.5\%$.

\section{SO(6) Matrix part and Young Tableaux}
\label{Young tableaux}
\setcounter{equation}{0}

This brief section is devoted to the matrix structure of scalars. It encodes the $SO(6)$ symmetry and does not depend on the coupling constant, therefore it is a quite general object, to some extent, independent of the theory (with specific symmetry) and the operator. Besides, many considerations can be repeated for other groups. 
In concrete, we will show that a systematic evaluation of $\Pi_{mat}^{(2n)}$ (\ref{Pi_mat}) by residues is possible, despite its cumbersome appearance.  Inspiration for this has been borrowed from the random partition method of $\mathcal{N}=2$ SYM theories \cite{Nekrasov}. By residues, we can perform the integrals over the auxiliary variables $a$,$c$ and obtain
\be\label{Pi_mat2-bis}
\Pi_{mat}^{(2n)}(u_1,\cdots,u_{2n})=\frac{4n^2}{(2n)!(n!)^2}\int \displaystyle\prod_{i=1}^{2n}\frac{db_i}{
2\pi}\frac{[\delta_{2n}(b_1,\dots,b_{2n})]^2}{\displaystyle\prod_{i,j}^{2n}f(u_i-b_j)}\displaystyle\prod_{i<j}\frac{b_{ij}^2}{(b_{ij}^2+1)} \, ,
\ee
where $\delta_{2n}$ is a known (though involved) symmetric polynomial depending only on the $b_i$ differences \cite{35}. This formula shares many similarities with the integral representation of the $2n$-instanton contribution to the Nekrasov partition function in a $\mathcal{N}=2$ $U(2n)$ SYM theory: we may loosely relate the rapidities $u_i$ to the $U(2n)$ scalar VEVs and the $b_i$ to the instanton coordinates.
In practice, it is well known that a representation as a sum over Young Tableaux configurations exists for the Nekrasov partition functions \cite{Nekrasov}, due to the particular pattern of poles and zeroes of the integrand: a very similar pattern appears here and motivates the following outcome. The main observation is that many residue configurations give the same contribution or are related to each other by permutation; so that we can write
\be
\Pi_{mat}^{(2n)}(u_1,\cdots,u_{2n})=\sum_{l_1+\cdots +l_{2n}=2n, l_i<3, l_i\geq l_{i+1}}(l_1,\cdots,l_{2n})_s= \sum_{|Y|=2n, l_i<3} (Y)_s \, ,\ee
where the residues in $b_j$ are encoded in the Young tableaux ($l_j\geq l_{j+1}$) $Y=(l_1,\cdots,l_{2n})$ with $l_j$ piled boxes (rows)\footnote{The constraint $l_j<3$ comes from the properties of $\delta_{2n}$.} at the $j=1,...,2n$ column. Yet, to complete the sum we need to add all the permutations of columns (a vector of Young columns without the property $l_j\geq l_{j+1}$), {\it i.e.} all the symmetric contributions in the rapidities $u_i$
\be
(l_1,\cdots,l_{2n})_s=(l_1,\cdots,l_{2n}) + \textit{perm.}.
\ee
In detail, the vector $(l_1,\cdots,l_{2n})$ represent a by-residue evaluation at $2n$ integrand poles $b_j$ (endowed with a multiplicity factor $(2n)!$ for permutation of the integration variables), with $l_j$ of them starting from $u_j+i/2$ and displaced by $i$ at the $j$-th column\footnote{So to say, any column $j$ is associated to the real rapidity $u_j$.}. The first examples should clarify: for $n=1$ we produce
\be
\Pi_{mat}^{(2)}(u_1,u_2)=(2,0)_s + (1,1)_s
\ee
where
\ba
(1,1)_s&=&(1,1)=\frac{4}{\left[(u_1-u_2)^2+1\right]^2}            \nonumber \\
(2,0)_s=(2,0)+(0,2)&=& \frac{1}{(u_1-u_2)(u_1-u_2+i)^2(u_1-u_2+2i)} + \{u_1\leftrightarrow u_2\}\, .
\ea
In $(1,1)$ one residue is taken in $b_1=u_1+i/2$ and the other in $b_2=u_2+i/2$ (times $2!$ permutations), while for $(2,0)$ in $b_1=u_1+i/2$ and $b_2=u_1+3i/2$ (times $2!$). The first non-trivial case is $n=2$ with four particles 
\be
\Pi_{mat}^{(4)}=(1,1,1,1)_s+(2,1,1,0)_s+(2,2,0,0)_s
\ee
with
\ba\label{Young4}
(1,1,1,1)&=&   4\left[\delta_4(u_1,u_2,u_3,u_4)\right]^2\displaystyle\prod_{i<j}^{4}\frac{1}{(u_{ij}^2+1)^2}\nonumber \\
(2,1,1,0)&=&\left[\delta_4(u_1,u_1+i,u_2,u_3)\right]^2\displaystyle\prod_{i<j}^3\frac{1}{(u_{ij}^2+1)^2}\frac{(u_{12}+i)^2}{(u_{12}+i)^2+1}\frac{(u_{13}+i)^2}{(u_{13}+i)^2+1}\times \nn\\
&& \frac{1}{\displaystyle\prod_{j=2}^{4}(u_1-u_j+i)(u_1-u_j+2i)\prod_{j=1}^3(u_j-u_4)(u_j-u_4+i)}      \nonumber \\
(2,2,0,0)&=& \frac{1}{\displaystyle\prod_{i=1}^2\prod_{j=3}^4(u_i-u_j)(u_i-u_j+i)^2(u_i-u_j+2i)}
\ea
where $\delta_4$ is a known symmetric polynomial depending on differences; the other contributions are obtained from (\ref{Young4}) by permutations of $u_i$ ({\it e.g.} $(2,2,0,0)_s=(2,2,0,0) + (2,0,2,0) +(2,0,0,2) +(0,2,2,0) +(0,2,0,2) +(0,0,2,2)$). An explicit formula for any $(l_1,\cdots,l_{2n})$ with more details will be given in a forthcoming publication \cite{35}. There we will also show how this educated guess of the polar structure ($P_{2n}$ is a polynomial, partially fixed by (\ref{factorization}))
\be
\label{P2n}
\Pi_{mat}^{(2n)}=\frac{P_{2n}(u_1,\cdots,u_{2n})}{\displaystyle\prod_{i<j}^{2n}(u_{ij}^2+1)(u_{ij}^2+4)}
\ee
can be proven by means of the factorisation (\ref{factorization}). For the time being, we need to highlight how this explicit method of computation is much useful also for the calculation of the connected functions.

\medskip
{\bf Acknowledgements} We thank J-E. Bourgine, I. Kostov and D. Serban for discussions. This project was  partially supported by the grants: GAST (INFN), UniTo-SanPaolo Nr TO-Call3-2012-0088, the ESF Network HoloGrav (09-RNP-092 (PESC)), the MPNS--COST Action MP1210 and the EC Network Gatis. AB acknowledges CEA and the Institut de Physique Th\'eorique (IPhT) for financial support and hospitality.


\end{document}